# Influence of Speech Codecs Selection on Transcoding Steganography

Artur Janicki, Wojciech Mazurczyk, Krzysztof Szczypiorski

Warsaw University of Technology, Institute of Telecommunications

Warsaw, Poland, 00-665, Nowowiejska 15/19

**Abstract.** The typical approach to steganography is to compress the covert data in order to limit its size, which is reasonable in the context of a limited steganographic bandwidth. TranSteg (Trancoding Steganography) is a new IP telephony steganographic method that was recently proposed that offers high steganographic bandwidth while retaining good voice quality. In TranSteg, compression of the overt data is used to make space for the steganogram. In this paper we focus on analyzing the influence of the selection of speech codecs on hidden transmission performance, that is, which codecs would be the most advantageous ones for TranSteg. Therefore, by considering the codecs which are currently most popular for IP telephony we aim to find out which codecs should be chosen for transcoding to minimize the negative influence on voice quality while maximizing the obtained steganographic bandwidth.

**Key words:** IP telephony, network steganography, TranSteg, information hiding, speech coding

## 1. Introduction

Steganography is an ancient art that encompasses various information hiding techniques, whose aim is to embed a secret message (steganogram) into a carrier of this message. Steganographic methods are aimed at hiding the very existence of the communication, and therefore any third-party observers should remain unaware of the presence of the steganographic exchange. Steganographic carriers have evolved throughout the ages and are related to the evolution of the methods of communication between people. Thus, it is not surprising that currently telecommunication networks are a natural target for steganography. The type of modern steganography that utilizes network protocols and/or relationships between them as the carrier for steganograms to enable hidden communication is called *network steganography*.

IP telephony is one of the most important services in the IP world and is changing the entire telecommunications landscape. It is a real-time service which enables users to make phone calls through IP data networks. An IP telephony connection consists of two phases, in which certain types of traffic are exchanged between the calling parties: signaling and conversation phases. In the first phase certain signaling protocol messages, for example SIP (Session Initiation Protocol) messages [27], are exchanged between the caller and callee. These messages are intended to set up and negotiate the connection parameters between the calling parties. In the latter phase two audio streams are sent bidirectionally. RTP (Real-Time Transport Protocol) [29] is most often utilized for voice data transport, and thus packets that carry voice payload are called RTP packets.

The consecutive RTP packets form an RTP stream. Due to the popularity of IP telephony, as well as the large volume of data and the variety of protocols involved, it is currently attracting the attention of the research community as a perfect carrier for steganographic purposes [18].

TranSteg (Transcoding Steganography) is a new steganographic method that has been introduced recently in [39]. It is intended for a broad class of multimedia and real-time applications, but its main foreseen application is IP telephony. TranSteg can also be exploited in other applications or services (like video streaming), wherever a possibility exists to efficiently compress (in a lossy or lossless manner) the overt data. The typical approach to steganography is to compress the *covert data* in order to limit its size (it is reasonable in the context of a limited steganographic bandwidth). In TranSteg, compression of the *overt data* is used to make space for the steganogram. TranSteg is based on the general idea of transcoding (lossy compression) of the voice data from a higher bit rate codec (and thus greater voice payload size) to a lower bit rate codec (with smaller voice payload size) with the least possible degradation in voice quality.

In [39] a proof of concept implementation of TranSteg was subjected to experimental evaluation to verify whether it is feasible. The obtained experimental results proved that it offers a high steganographic bandwidth (up to 32 kbit/s) while introducing delays lower than 1 ms and still retaining good voice quality.

In this paper we focus on analyzing how the selection of speech codecs impacts hidden transmission performance, that is, which codecs would be the most advantageous ones for TranSteg. Therefore, the main contribution of the paper is to establish, by considering the codecs for IP telephony which are currently most popular, which speech codecs should be chosen for transcoding to minimize the negative influence on the hidden data carrier (voice quality) while maximizing the obtained steganographic bandwidth.

The rest of the paper is structured as follows. Section 2 presents related work on IP telephony steganography. Section 3 describes the functioning of TranSteg and its hidden communication scenarios. Section 4 presents the experimental methodology and results obtained. Finally, Section 5 concludes our work.

## 2. Related Work

IP telephony as a hidden data carrier can be considered a fairly recent discovery. The proposed steganographic methods stem from two distinctive research origins. The first is the well-established image and audio file steganography [6], which has given rise to methods which target the digital representation of voice as the carrier for hidden data. The second sphere of influence is the so-called covert channels, created in different network protocols [1], [24] (a good survey on covert channels, by Zander et al., can be found in [37]); these solutions target specific VoIP protocol fields (e.g. signaling protocol – SIP, transport protocol – RTP, or control protocol – RTCP) or their behavior. Presently, steganographic methods that can be utilized in telecommunication networks are jointly described by the term *network steganography*, or, specifically, when applied to IP telephony, by the term *steganophony* [18].

The first VoIP steganographic methods to utilize the digital voice signal as a hidden data carrier were proposed by Dittmann et al. in 2005 [9]. The authors evaluated the existing audio steganography techniques, with a special focus on the solutions which were suitable for IP telephony. This work was further extended and published in 2006 in [17]. In [40], an implementation of SteganRTP was described. This tool employed the least significant bits (LSB) of the G.711 codec to carry steganograms. Wang and Wu, in [34], also suggested using the least significant bits of voice samples to carry secret communication, but here the bits of the steganogram were coded using a low rate voice codec, like Speex. In [31], Takahashi and Lee proposed a similar approach and presented its proof of concept implementation, Voice over VoIP ($Vo^2IP$), which can establish a hidden conversation by embedding compressed voice data into the regular, PCM-based, voice traffic. The authors also considered other methods that can be utilized in VoIP steganography, like DSSS (Direct Sequence Spread Spectrum), FHSS (Frequency-Hopping Spread Spectrum), or Echo hiding. In [3], Aoki proposed a steganographic method based on the characteristics of PCMU (Pulse Code Modulation) in which the zeroth speech sample can be represented by two codes due to the overlap. Another LSB-based method was proposed by Tian et al. in [33]. The authors incorporated the m-sequence technique to eliminate the correlation among secret messages to resist statistical detection. A similar approach, also LSB-based, relying on adaptive VoIP steganography was presented by the same authors in [32]; a proof of concept tool, StegTalk, was also developed. In [36] Xu and Yang proposed an LSB-based method dedicated to voice transmission using the G.723.1 codec in 5.3 kbps mode. They identified five least significant bits in various G.723.1 parameters and used them to transmit hidden data; the method provided a steganographic bandwidth of 133.3 bps. In [23] Miao and Huang presented an adaptive steganography scheme based on the smoothness of the speech block. Such an approach proved to give better results in terms of voice quality than the LSB-based method. An interesting study is described in [26], where Nishimura proposed hiding information in the AMR-coded stream by using an extended quantization-based method of pitch delay (one of the AMR codec parameters). This additional data transmission channel was used to extend the audio bandwidth from narrow-band (0.3–3.4 kHz) to wide-band (0.3–7.5 kHz).

Utilization of the VoIP-specific protocols as a steganogram carrier was first proposed by Mazurczyk and Kotulski in 2006 [19]. The authors proposed using covert channels and watermarking to embed control information (expressed as different parameters) into VoIP streams. The unused bits in the header fields of IP, UDP, and RTP protocols were utilized to carry the type of parameter, while the actual parameter value is embedded as a watermark into the voice data. The parameters are used to bind control information, including data authentication, to the current VoIP data flow. In [21] and [22] Mazurczyk and Szczypiorski described network steganography methods that can be applied to VoIP: to its signaling protocol, SIP (with SDP), and to its RTP streams (also with RTCP). They discovered that a combination of information hiding solutions provides a capacity to covertly transfer about 2000 bits during the signaling phase of a connection and about 2.5 kbit/s during the conversation phase. In [22], a novel method called LACK (Lost Audio Packets Steganography) was introduced; it was later described and analyzed in [20] and [18]. LACK relies on the modification of both the content of the RTP packets and their time dependencies. This method takes advantage of the fact that, in typical multimedia communication protocols like RTP, excessively delayed packets are not used for the reconstruction of the transmitted data at the receiver; that is, the packets are considered useless and discarded. Thus, hidden communication is possible by introducing intentional delays into selected RTP packets and substituting the original payload with a steganogram.

Bai et al. [4] proposed a covert channel based on the jitter field of the RTCP header. This is performed in two stages: firstly, statistics of the value of the jitter field in the current network are calculated. Then, the secret message is modulated into the jitter field according to the previously calculated parameters. Utilization of such modulation guarantees that the characteristic of the covert channel is similar to that of the overt one. In [8], Forbes proposed a new RTP-based steganographic method that modifies the timestamp value of the RTP header to send steganograms. The method's theoretical maximum steganographic bandwidth is 350 bit/s.

The TranSteg technique that was first introduced in [39] is a development of the last of the discussed groups of steganographic methods for VoIP, originating from covert channels. Compared to the existing solutions, its main advantages are a high steganographic bandwidth, low steganographic cost (i.e. little degradation of voice quality), and difficult detection.

## 3. TranSteg Functioning

TranSteg, like every steganographic method, can be described by the following set of characteristics: its steganographic bandwidth, its undetectability, and the steganographic cost. The term "steganographic bandwidth" refers to the amount of secret data that can be sent per time unit when using a particular method. Undetectability is defined as the inability to detect a steganogram within a certain carrier. The most popular way to detect a steganogram is to analyze statistical properties of the captured data and compare them with the typical values for that carrier. Lastly, the steganographic cost characterizes the degradation of the carrier caused by the application of the steganographic method. In the case of TranSteg, this cost can be expressed by providing a measure of the conversation quality degradation induced by transcoding and the introduction of an additional delay.

The general idea behind TranSteg is as follows (Fig. 1). RTP packets carrying user's voice are inspected and the codec originally used for speech encoding (here called the *overt codec*) is determined by analyzing the PT (Payload Type) field in the RTP header (Fig. 1, 1). If typical transcoding occurs then the original voice frames are usually recoded using a different speech codec to achieve a smaller voice frame (Fig. 1, 2). But in TranSteg an appropriate *covert codec* for the overt codec used originally is selected. The application of the *covert codec* yields a comparable voice quality but a smaller voice payload size than originally. Next, the voice stream is transcoded, but the original, large voice payload size and the codec type indicator are preserved, and thus the PT field is left unchanged. Instead, after placing the transcoded voice of a smaller size inside the original payload field, the remaining free space is filled with hidden data (Fig. 1, 3). Of course, the steganogram does not necessarily need to be inserted at the end of the payload field. It can be spread across this field or mixed with voice data as well. We assume that for the purposes of this paper it is not crucial which steganogram spreading mechanism is used, and thus it is out of the scope of this work.

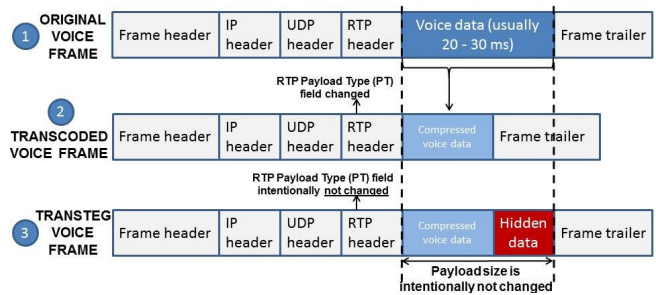

**Fig. 1:** Frame bearing voice payload encoded with overt codec (1), typically transcoded (2), and encoded with covert codec (3)

The performance of TranSteg depends, most notably, on the characteristics of the pair of codecs: the *overt codec* originally used to encode user speech and the *covert codec* utilized for transcoding. In ideal conditions the covert codec should:

- not significantly degrade user voice quality when compared to the quality of the overt codec (in an ideal situation there should be no negative influence at all),
- provide the smallest achievable voice payload size, as this results in the most free space in an RTP packet to convey a steganogram.

On the other hand, the overt codec in an ideal situation should:

- result in the largest possible voice payload size to provide, together with the covert codec, the highest achievable steganographic bandwidth,
- be commonly used, to avoid arousing suspicion.

TranSteg can be utilized in four hidden communication scenarios (Fig. 2). The first scenario (S1 in Fig. 2) is the most common and typically the most desired: the sender and the receiver conduct a VoIP conversation while simultaneously exchanging steganograms (end-to-end). The conversation path is identical to the hidden data path. In the next three scenarios (marked S2–S4 in Fig. 2) only a part of the VoIP end-to-end path is used for hidden communication. As a result of actions undertaken by intermediate nodes, the sender and/or the receiver are, in principle, unaware of the steganographic data exchange. The application of TranSteg in IP telephony connections offers a chance to preserve users' conversation and simultaneously transfer steganograms. As noted previously, this is especially important for scenarios S2–S4.

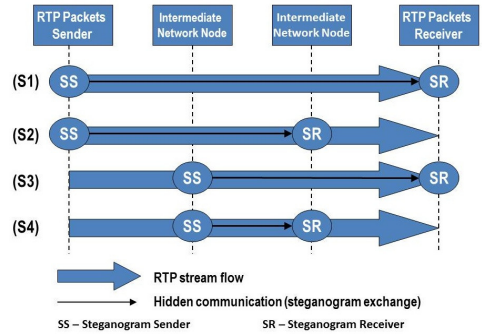

**Fig. 2:** Hidden communication scenarios for TranSteg

In this paper, we consider TranSteg in scenario S4 because it is the worst case scenario in terms of the speech quality, as it requires triple transcoding (and two transcodings result from the TranSteg functioning). If TranSteg scenarios S1–S3 were applied, we would avoid one or even two (in scenario S1) transcodings, and therefore the negative influence on speech quality would be lower than presented in this study.

In scenario S4 (Fig. 3) it is assumed that both SS and SR are able to intercept and analyze all RTP packets exchanged between the calling parties. Neither SS nor SR is involved in the IP telephony conversation as an overt calling party. Thus, it is harder to detect hidden communication between the SS and SR compared to the previously described scenarios (since neither is an initiator of the overt traffic).

In the presented scenario the behavior of SS and SR is similar: they both perform transcoding, SS from overt to covert, and SR from covert to overt codecs. The steganogram is exchanged only along the part of the communication path where the RTP stream travels "inside" the network – it never reaches the endpoints. The steps of the TranSteg scenario for SS are as follows:

- **Step 1:** For an incoming RTP stream it transcodes the user's voice data from the overt to the covert codec.
- **Step 2:** The transcoded voice payload is placed once again in an RTP packet. The RTP packet's header remains unchanged.
- **Step 3:** The remaining free space of the RTP payload field is filled with the steganogram's bits (and thus the original voice payload is erased).
- **Step 4:** Checksums in lower layer protocols (UDP checksum and CRC at the data link) are adjusted.
- **Step 5:** Modified frames with encapsulated RTP packets are sent to the receiver (SR).

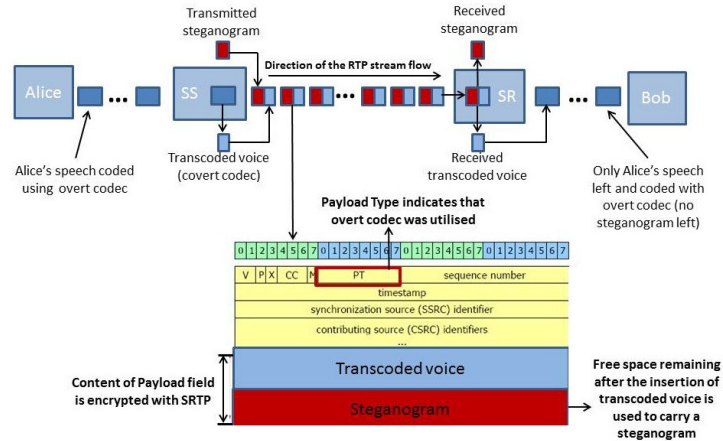

**Fig. 3:** The TranSteg concept, scenario S4 (SS – Steganogram Sender; SR – Steganogram Receiver)

Next, when the altered RTP stream reaches the SR, it performs the following steps:

- **Step 1:** It extracts the voice payload and the steganogram from the RTP packets.
- **Step 2:** The voice payload is then transcoded from the covert to the overt codec and placed once again in consecutive RTP packets. By performing this task the steganogram is overwritten with user voice data. The RTP packet's header remains unchanged.
- **Step 3:** Checksums for the lower layer protocols (i.e. the UDP checksum and CRC at the data link, if they have been utilized) are adjusted.
- **Step 4:** Modified frames with encapsulated RTP packets are sent to the receiver (callee).

The most significant advantage of scenario S4 from Fig. 2 is its potential use of aggregated IP telephony traffic to transfer steganograms. If both SS and SR have access to more than one VoIP call then the achievable steganographic bandwidth can be significantly increased.

The SS and SR have limited influence on the choice of the overt codec, because they are both located at some intermediate network node. Due to this fact they are bound to rely on the codec chosen by the overt, non-steganographic calling parties or they can interfere with the choice of the overt codec during the signaling phase of the call where the codec negotiation is taking place. When relying on the first option, SS and SR must be able to choose the covert codec in such a way as to maximize the achievable steganographic bandwidth while minimizing the steganographic cost.

This paper focuses on analyzing the best covert codec choices for the speech codecs that are currently the most popular ones utilized for IP telephony (overt codecs) in terms of steganographic bandwidth and cost.

## 4. TranSteg Experimental Results

### 4.1 Experiment Methodology

In our experiments we emulated 20 unidirectional voice transmission channels. We took the information about location of speech activity (turn-taking patterns) and background noise from the LUNA corpus [25], containing real phone conversations between travelers and a public transport information line. Voice activity ranged from 40.5% to 67.5%, with an average of 46.5%. The speech signal was taken from the TSPSpeech corpus [16] for English and CORPORA [11] for Polish. Each of these databases contains phonetically balanced sentences in the respective languages. In such a way we generated 20 one-minute recordings, 10 in English and 10 in Polish, sampled at 8 kHz with 16-bit resolution. Each language group consisted of five male and five female speakers.

As the overt codec we used four of the most popular codecs used in the Internet telephony [41], [42]:

- G.711: a codec designed originally for fixed telephony [13], but also used in VoIP due to its simplicity and good speech quality; it is just an implementation of logarithmic quantization with 8 bits per sample, thus offering a bitrate of 64 kbps. The option *A*-law, which is used most in the world, was researched in this study.
- Speex: a CELP (Code Excited Linear Prediction)-based lossy codec designed specifically for VoIP applications [35]. Although it allows wide-band and ultra-band transmissions, only the narrow-band variant was considered here. It offers 10 different compression levels corresponding to 10 different bitrates, of which three modes were selected: (i) mode 7, the highest mode designed for speech, working with a bitrate of 24.6 kbps, hereinafter called Speex(7), (ii) a moderate mode 4, requiring a bitrate of 11.0 kbps, hereinafter called Speex(4), and (iii) mode 2, which is the lowest recommended mode for speech, working at 5.95 kbps, called here Speex(2).
- iLBC: another low-bitrate CELP-based codec designed for VoIP, using frame-independent long term prediction, thus making it resistant to packet losses [2]. Depending on the analysis frame length (20 ms or 30 ms), it requires 15.2 kbps or 13.33 kbps, respectively. Twenty-millisecond frames were used in this study.
- G.723.1: a codec based on MP-MLQ (Multi-Pulse Maximum Likelihood Quantization) and ACELP (Algebraic CELP), offering bitrates of 5.3 kbps and 6.4 kbps, respectively. In this study the latter option was used.

As for the covert codecs, apart from the abovementioned ones, we tested the following:

- G.711.0: also known as G.711 LLC (LossLess Compression), is a lossless extension of the G.711 codec, standardized by ITU fairly recently in 2009 [12]. It works with various frame lengths (40–320 samples); however in this study 160-sample (20 ms) frames were used. Due to its losslessness it offers a variable bitrate, as the compression ratio depends on the actual voice data. It is also stateless, which means that the encoding of a particular frame does not depend on the previous or the next frame, making it suitable for packet transmission, including the TranSteg technique.

- G.726: an ADPCM (Adaptive Differential PCM) codec, standardized by ITU-T in 1990 [7], offering bitrates from 16 kbps up to 40 kbps. In this study we used the most common option, 32 kbps, which was already tried with TranSteg in [39].
- GSM 06.10 [10] (also known as GSM Full-Rate or GSM-FR): designed in the early 1990s for the GSM telephony, but used in VoIP as well. It is based on the RPE-LTP (Regular Pulse Excitation – Long Term Prediction) algorithm, with the use of the LPC technique.
- AMR (Adaptive Multi-Rate): a codec adopted in 1999 as standard by 3GPP, used widely in GSM and UMTS [28]. It is based on CELP, but also incorporates other techniques, such as DTX (Discontinuous Transmission) and CNG (Comfort Noise Generation). It covers eight different bitrates, from 4.75 kbps up to 12.2 kbps. The highest 12.2 kbps mode, used further in this study, is compatible with ETSI GSM EFR (Enhanced Full-Rate).
- G.729: operates at a bitrate of 8 kbps, and is based on CS-ACELP (Conjugate Structure ACELP). Several annexes to the basic G.729 have been published so far. In this study we used Annex A, which has slightly lower computational requirements than the original G.729.

To sum up, we tested codecs from various families: waveform codecs, that is, the ones preserving the speech signal shape (G.711, G.726), CELP-based codes, that is, the ones based on linear prediction excited by vectors from a codebook (Speex, iLBC, G.723.1, AMR), an RPE-LTP codec (GSM 06.10) and, last but not least, a lossless audio codec (G.711.0).

Emulations were conducted in the Matlab® 7.12 environment. The codecs' functionality was implemented using the SoX toolbox version 14.3.2 [30], the G.723.1 Speech Coder and Decoder MATLAB toolbox, and reference implementations provided by ITU-T and iLBCfreeware.org.

Speech quality was assessed using the PESQ algorithm [14], by comparing the original and output audio files (see Fig. 4). As a reference, first the single transcoding was emulated (configuration A). Later, the double one was tried, but still without the covert transcoding (configuration B). The experiments with the actual choice of overt and covert codecs were run in configuration C, which emulates TranSteg operation in scenario S4. This scenario requires triple transcoding. The PESQ algorithm returned MOS-LQO (Mean Opinion Score – Listening Quality Objective) results, which were further averaged over all the tested signals. In addition, the confidence interval at the confidence level of 95% was calculated.

Additionally, for the lossless covert codec G.711.0 we measured the bitrate, as it is a variable bitrate codec. We assumed that in configurations with G.711.0 as the covert codec, one byte in the payload field will be used for signaling to inform how many bytes in a given packet are used for the overt transmission.

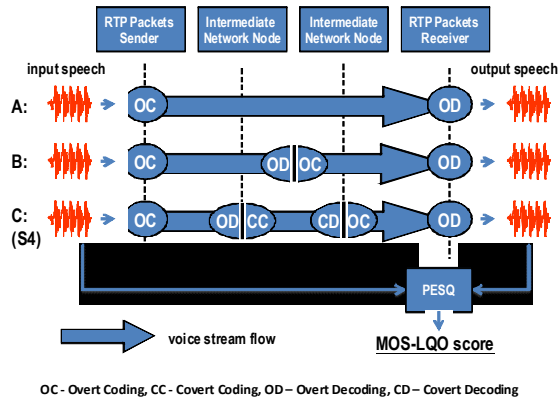

**Fig. 4:** The experimental setup

### 4.2 Experimental Results

### 4.2.1 Results for Steganographic Bandwidth

We calculated the steganographic bandwidth for all pairs of overt and covert codecs considered. The results, presented in

Table I, are based on the bitrate required by the overt and covert codecs. The result for the G.711/G.711.0 (overt/covert) pair was obtained experimentally by measuring the bitrate of the lossless G.711.0. For the 20 tested signals it ranged between 22.77 kbps and 37.83 kbps, giving an average of 31.11 kbps, with a standard deviation of 3.67 kbps. The steganographic bandwidth in this case takes into account the signaling bits (8 bits per 20-ms frame).

Table I. Steganographic bandwidth [kbps] for various sets of overt and covert codecs (unfeasible combinations are grayed out). An asterisk (*) denotes the result with a confidence interval of ±2.27 kbps at the 95% confidence level. The top three results are bolded.

| | | | G.711 | Speex(7) | iLBC | Speex(4) | G.723.1 | Speex(2) |
|---|---|---|---|---|---|---|---|---|
| Covert codec | G.711 | 64 | | | | | | |
| | G.711.0 | 32.49* | 31.11* | | | | | |
| | G.726 | 32 | 32 | | | | | |
| | Speex(7) | 24.6 | 39.4 | | | | | |
| | iLBC | 15.2 | 48.8 | 9.4 | | | | |
| | GSM 06.10 | 13 | 51 | 11.6 | 2.2 | | | |
| | AMR | 12.2 | 51.8 | 12.4 | 3 | | | |
| | Speex(4) | 11 | 53 | 13.6 | 4.2 | | | |
| | G.729 | 8 | **56** | 16.6 | 7.2 | 3 | | |
| | G.723.1 | 6.3 | **57.7** | 18.3 | 8.9 | 4.7 | | |
| | Speex(2) | 5.95 | **58.05** | 18.65 | 9.25 | 5.05 | 0.35 | |
| | Bitrate [kbps] | | 64 | 24.6 | 15.2 | 11 | 6.3 | 5.95 |
| | Codec | | G.711 | Speex(7) | iLBC | Speex(4) | G.723.1 | Speex(2) |
| | | | Overt codec | | | | | |

The pairs in which the covert codec required a higher bandwidth than the overt one were found to be unfeasible in the TranSteg technique and were therefore excluded from further experiments (in Table I they are grayed out). It is worth noting that the steganographic bandwidth depends strongly on the codec used in the overt channel – it is the highest for G.711, ranging from 32 kbps up to 58.08 kbps, whilst it is the lowest for G.723.1, Speex(4), and iLBC, allowing only a few kilobits per second. When the overt transmission uses Speex(2), steganographic transmission using the TranSteg technique is not possible at all, due to the low overt transmission bitrate (5.95 kbps). Considering only the bitrate, it turned out that the codecs G.711.0, G.726, and Speex(7) can serve as covert codecs only when the overt voice transmission is using G.711.

### 4.2.2 Results for Steganographic Cost

The TranSteg technique requires the speech signal to be transcoded twice with the overt codec. The experiments were run to verify how much single and double transcoding (configurations A and B in Fig. 4) impacts the voice transmission quality without utilizing a covert codec yet. The results in Table II show that G.711 was the most resistant against double transcoding. This is due to the fact that G.711 is a waveform codec. All the remaining, CELP-based codecs yielded decreases in quality ranging from 0.24 MOS for iLBC to 0.51 MOS for Speex(4).

Table II. Initial voice transmission quality [MOS] – no TranSteg used

| Codec | G.711 | Speex(7) | iLBC | Speex(4) | G.723.1 |
|---|---|---|---|---|---|
| Single transcoding | 4.46 | 4.17 | 3.92 | 3.62 | 3.70 |
| Double transcoding | 4.46 | 3.92 | 3.68 | 3.11 | 3.24 |

Table III presents the results of overall voice quality of the output speech signal for all the tested configurations with emulated TranSteg (configuration C in Fig. 4). It shows that most of the pairs retained an acceptable level of quality – more than 3 in the MOS scale; some pairs with G.711 as the overt codec yielded results even higher than 4, which is considered "more than good". However, the configurations with G.723.1 and Speex(4) as the overt codec provided voice quality lower than 3, which was found to be unacceptably low. It is noteworthy that the overall quality obtained with the G.711/G.711.0 configuration is the same as for G.711 without TranSteg, as shown in Table II.

Table III. Overall voice quality [MOS] for various sets of overt and covert codecs (unfeasible combinations are grayed out). The best results (greater than 4 MOS) are highlighted; the worst ones (less than 3 MOS) are italicized.

| Covert codec | G.711 | | | | | |
|---|---|---|---|---|---|---|
| | G.711.0 | **4.46** | | | | |
| | G.726 | **4.04** | | | | |
| | Speex(7) | **4.11** | | | | |
| | iLBC | 3.88 | 3.66 | | | |
| | GSM 06.10 | 3.60 | 3.41 | 3.34 | | |
| | AMR | **4.10** | 3.73 | 3.46 | | |
| | Speex(4) | 3.59 | 3.27 | 3.28 | | |
| | G.729 | 3.72 | 3.43 | 3.26 | *2.76* | |
| | G.723.1 | 3.65 | 3.43 | 3.29 | *2.80* | |
| | Speex(2) | 3.27 | 3.04 | 3.08 | *2.57* | *2.71* |
| | Codec | G.711 | Speex(7) | iLBC | Speex(4) | G.723.1 |
| | | Overt codec | | | | |

Having measured the overall quality with TranSteg and compared it with the initial values, we were able to calculate the steganographic cost introduced by various configurations. The results are shown in Table IV. The cost for G.711/G.711.0 was obviously equal to zero, due to the lossless nature of the covert codec. When lossy codecs were used as covert ones, the steganographic cost ranged from as low as around 0.35 MOS for G.711/Speex(7) and G.711/AMR pairs up to around 1.2 MOS for G.711/Speex(2) and Speex(7)/Speex(2) pairs. The confidence interval for the 20 tested signals, assuming a 95% confidence level, turned out to be acceptably narrow (see Table IV), being usually far below ± 0.1 MOS (with only one exception, for G.711/GSM 06.10, where it yielded ± 0.104).

Table IV. Results of steganographic cost [MOS] for the tested variants, with confidence intervals (at 95% confidence level). The best results (less than 0.5 MOS) are highlighted; the worst ones (greater than 1 MOS) are italicized.

| Covert codec | G.711 | | | | | |
|---|---|---|---|---|---|---|
| | G.711.0 | **0.00 ± 0.000** | | | | |
| | G.726 | **0.42 ± 0.067** | | | | |
| | Speex(7) | **0.35 ± 0.052** | | | | |
| | iLBC | 0.59 ± 0.052 | 0.50 ± 0.055 | | | |
| | GSM 06.10 | 0.86 ± 0.104 | 0.76 ± 0.086 | 0.58 ± 0.079 | | |
| | AMR | **0.36 ± 0.039** | **0.43 ± 0.059** | **0.46 ± 0.069** | | |
| | Speex(4) | 0.88 ± 0.069 | 0.90 ± 0.064 | 0.64 ± 0.074 | | |
| | G.729 | 0.74 ± 0.066 | 0.74 ± 0.074 | 0.66 ± 0.070 | 0.86 ± 0.088 | |
| | G.723.1 | 0.81 ± 0.059 | 0.74 ± 0.065 | 0.63 ± 0.074 | 0.82 ± 0.089 | |
| | Speex(2) | *1.20 ± 0.067* | *1.13 ± 0.074* | *0.84 ± 0.083* | *1.04 ± 0.074* | 0.99 ± 0.087 |
| | Codec | G.711 | Speex(7) | iLBC | Speex(4) | G.723.1 |
| | | Overt codec | | | | |

### 4.2.3 Steganographic Cost versus Steganographic Bandwidth

Figure 5 shows graphically the steganographic cost of various tested configurations against the offered steganographic bandwidth. It can be observed that, for a given overt codec, a choice of the most efficient covert codec in the TranSteg technique is often (but not always) a matter of a trade-off between the steganographic bandwidth and the steganographic cost. For example, with iLBC in the overt channel, in the covert channel we can use AMR, offering 3 kbps steganographic bandwidth with only 0.43 MOS of steganographic cost. If we choose Speex(2), TranSteg will create a 9.25 kbps steganographic channel, but at the expense of a higher cost: 0.84 MOS.

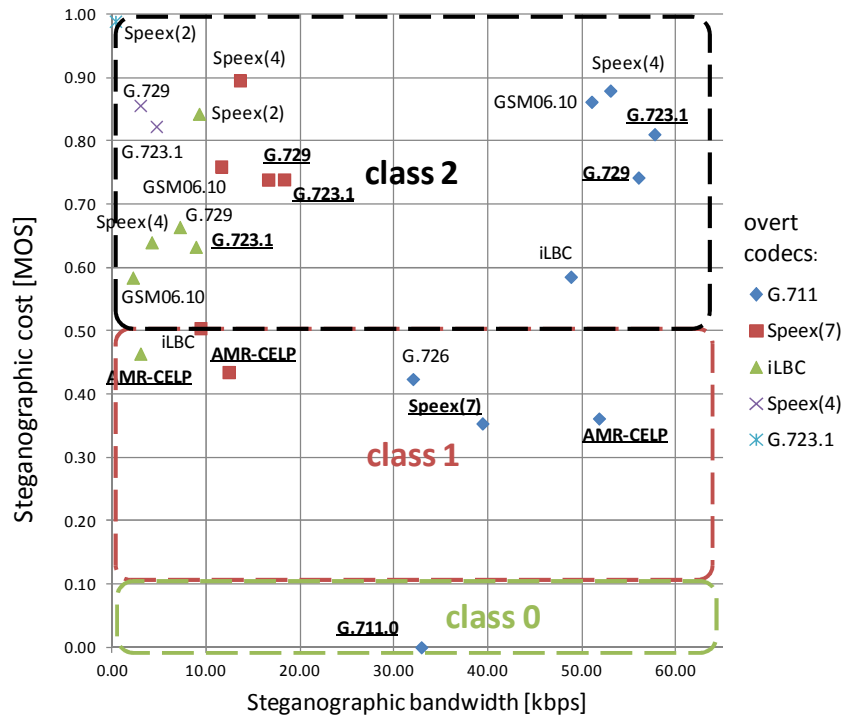

**Fig. 5:** Steganographic cost against the steganographic bandwidth for the tested overt/covert codec pairs. The labels inform about the covert codec.

As the decrease in the acceptable level of voice quality depends on an actual application and user requirements, we took the liberty of dividing the results in Fig. 5 into three classes:

- Class 0: no quality decrease; for configurations with steganographic cost lower than 0.1 MOS;
- Class 1: minor quality decrease; for configurations with steganographic cost between 0.1 and 0.5 MOS;
- Class 2: moderate quality decrease; for configurations with steganographic cost between 0.5 and 1.0 MOS.

The configurations with costs above 1.0 MOS are outside the picture, as we consider such a major decrease of voice quality in TranSteg to be unacceptable.

It must be emphasized that a high steganographic bandwidth **does not always** imply a high steganographic cost. For example, with G.711 in the overt channel we can use either GSM 06.10, Speex(4), or AMR in the covert channel, creating in each case a similar, capacious (ca. 50 kbps) steganographic channel. But GSM 06.10 and Speex(4) will cause a decrease of more than 0.85 MOS in the voice quality, while, remarkably, AMR will introduce a steganographic cost of only 0.36 MOS.

The experiments we ran helped to identify which codecs would provide better quality while providing similar bandwidth (or provide higher bandwidth while assuring similar quality). As a result, the configurations which we recommend in each class are underlined in Fig. 5. For example, for all the tested overt codecs, AMR introduced a much lower decrease in quality than GSM 06.10, even though they provide similar bandwidths. For iLBC in the overt channel, the recommended covert codecs are AMR and G.723.1, whilst the codecs Speex(4) and G.729 would provide lower steganographic bandwidth at a similar steganographic cost. The pairs which provided overall quality lower than 3 MOS (i.e. the ones with Speex(4) and G.723.1 as the overt codecs) are not recommended.

In general, we recommend only one pair in Class 0 (G.711 with lossless G.711.0), four pairs in Class 1, and five pairs in Class 2.

## 5. Conclusions and Future Work

TranSteg is a new steganographic method dedicated to multimedia services like IP telephony. In this paper the analysis of the influence of the selection of speech codecs on the performance of TranSteg hidden transmission was presented. By considering the codecs which are currently most popular for IP telephony we wanted to find out which codecs should be chosen for transcoding to minimize the negative influence on hidden data carriers while maximizing the obtained steganographic bandwidth.

The obtained experimental results show that TranSteg is most effective when G.711 is used for the overt transmission. We think that this is caused by several factors:

- high G.711 bitrate, so there is more space for hidden data;
- high speech quality offered by G.711;
- G.711 performs well if transcoded more than once (see Table II), which we think is due to the fact that G.711 is a waveform codec; that is, it preserves the waveform shape;
- while being a waveform codec, G.711 behaves well if further transcoded with other codecs, especially CELP-based ones (AMR, Speex in mode 7).

Therefore, when it is possible to select an overt codec (e.g., by imposing it in codec negotiation in scenario S3), G.711 is the best option. In contrast, Speex in the low mode (4) and G.723.1 as the overt codecs turned out to provide too low overall quality when cascaded with any of the covert codecs, and therefore should be avoided in TranSteg in overt transmission.

When experimenting with various combinations of overt and covert codecs, we observed that some codecs do not complement each other well. For example, Speex in mode 7 works significantly better (in terms of voice quality) with AMR than with GSM 06.10, even though the two TranSteg configurations result in similar steganographic bandwidths. A similar phenomenon was observed in other research projects, for example in speaker recognition from coded speech, in situations where there was a mismatch between the codec used in voice transmission and the codec used to create speakers' models [15].

The choice of a covert codec depends on an actual application, or more precisely, on whether priority is given to higher steganographic bandwidth or better speech quality. We recommended 10 pairs of overt/covert codecs which can be used effectively in TranSteg in various conditions depending on the required steganographic bandwidth, allowed steganographic cost, and the codec used in the overt transmission. We grouped those pairs into three classes based on the steganographic cost. The pair G.711/G.711.0 is costless; nevertheless it offers a remarkably high steganographic bandwidth, on average more than 31 kbps. However caution must be taken, as the G.711.0 bitrate is variable and depends on an actual signal being transmitted in the overt channel.

Codec AMR working in 12.2 kbps mode proved to be very efficient as the covert codec in TranSteg. This is a low bitrate codec which does not significantly degrade the quality: the steganographic cost ranged between 0.36 and 0.46 MOS.

In this research we showed results for scenario S4, which is the worst case scenario in terms of the speech quality, as it requires triple transcoding. If TranSteg scenarios S1–S3 were applied, we would avoid one or even two (in scenario S1) transcodings, and therefore steganographic cost would be lower than presented in this study.

Future work will include the development of the TranSteg-capable softphone, which will include results related to speech codec selection presented in this paper. Moreover, effective and efficient TranSteg detection methods will be pursued.

## ACKNOWLEDGMENTS

This work was partially supported by the Polish Ministry of Science and Higher Education and Polish National Science Centre under grants: DEC-2011/01/D/ST7/05054

[4] Bai LY, Huang Y, Hou G, Xiao B (2008) Covert channels based on jitter field of the RTCP Header. In: Int Conf Intelligent Information Hiding and Multimedia Signal Processing

[5] Baugher M, Casner S, Frederick R, Jacobson V (2004) The secure real-time transport protocol (SRTP), RFC 3711, March, 2004

[6] Bender W, Gruhl D, Morimoto N, Lu A (1996) Techniques for data hiding. IBM. Syst J 35(3/4):313–336

[7] CCITT (1990) Recommendation G.726: 40, 32, 24, 16 kbps adaptive differential pulse code modulation (ADPCM)

[8] Forbes CR, (2009) A New covert channel over RTP. MSc thesis, Rochester Institute of Technology. https://ritdml.rit.edu/bitstream/handle/1850/12883/CForbesThesis8-21-2009.pdf?sequence=1

[9] Dittmann J, Hesse D, Hillert R (2005) Steganography and steganalysis in voice-over IP scenarios: operational aspects and first experiences with a new steganalysis tool set. In: Proc SPIE, Vol 5681, Security, Steganography, and Watermarking of Multimedia Contents VII, San Jose, pp 607–618

[10] ETSI (2000) Digital cellular telecommunications system (Phase 2+) (GSM); Full rate speech; Transcoding (GSM 06.10 version 8.1.1 Release 1999)

[11] Grocholewski S (1997) CORPORA – Speech database for Polish diphones. In: 5th Europ Conf Speech Communication and Technology Eurospeech '97, Rhodes, Greece

[12] ITU-T (2009) Recommendation G.711.0: Lossless compression of G.711 pulse code modulation

[13] ITU-T (1998) Recommendation G.711: Pulse code modulation (PCM) of voice frequencies

[14] ITU-T (2001) Recommendation P.862: Perceptual evaluation of speech quality (PESQ): An objective method for end-to-end speech quality assessment of narrow-band telephone networks and speech codecs

[15] Janicki A, Staroszczyk T (2011) Speaker recognition from coded speech using support vector machines. In: Habernal I, Matoušek V (eds) Proc 14th Int Conf Text, Speech and Dialogue (TSD 2011), Lecture Notes on Artificial Intelligence (LNAI) 6836, pp 291–298, Springer-Verlag

[16] Kabal P (2002) TSP speech database, Tech Rep, Department of Electrical & Computer Engineering, McGill University, Montreal, Quebec, Canada

[17] Krätzer C, Dittmann J, Vogel T, Hillert R (2006) Design and evaluation of steganography for Voice-over-IP. In: Proc IEEE Int Symp Circuits and Systems (ISCAS), Kos, Greece

[18] Lubacz J, Mazurczyk W, Szczypiorski K (2010) Voice over IP (invited paper). In: IEEE Spectrum, ISSN: 0018-9235, Feb, pp 40–45

[19] Mazurczyk W, Kotulski Z (2006) New security and control protocol for VoIP based on steganography and digital watermarking, In: Proc 5th Int Conf Computer Science – Research and Applications (IBIZA 2006), Poland, Kazimierz Dolny, Feb 9–11

[20] Mazurczyk W, Lubacz J (2010) LACK – a VoIP steganographic method. Telecommun Syst: Model Anal Des Manag 45(2/3) ISSN: 1018-4864 (print version), ISSN: 1572-9451 (electronic version)

[21] Mazurczyk W, Szczypiorski S (2008) Covert channels in SIP for VoIP signaling. In: Jahankhani H, Revett K, Palmer-Brown D (eds) ICGeS 2008 – Communications in Computer and Information Science (CCIS) 12, Springer Verlag Berlin Heidelberg, Proc 4th Int Conf Global E-security 2008, London, United Kingdom, June 23–25, pp 65–70

[22] Mazurczyk W, Szczypiorski S (2008) Steganography of VoIP Streams. In: Meersman R, Tari Z (eds) OTM 2008, Part II – Lecture Notes in Computer Science (LNCS) 5332, Springer-Verlag Berlin Heidelberg, Proc OnTheMove
16